\documentclass[11pt]{article}
\pdfoutput=1
\usepackage{jpub,amsmath,amssymb,hyperref}
\usepackage{bm}
\usepackage{amsmath}
\usepackage{amssymb}
\usepackage{youngtab}
\usepackage{bbold}
\usepackage{pgfplots,todonotes}
\usepackage{tikz}
\usepackage[utf8]{inputenc}
\usepackage{breqn}
\usepackage{aas_macros}
\usepackage{colortbl}
\usepackage{mathtools}
\usepackage{amstext}
\usepackage{graphicx,epsfig}
\usepackage{subcaption}
\usepackage{fancybox}
\usepackage{slashed}
\usepackage{afterpage}
\usepackage{color}
\usepackage{ulem}
\usepackage{enumitem}
\usepackage{dsfont}
\usepackage{tabu}
\usepackage{tikz}
\usepackage{booktabs}
\usepackage[numbers,sort&compress]{natbib}
\usepackage{cancel}
\usepackage{framed}
\usepackage{bbding}
\usepackage{wasysym}
\usepackage{multirow,hhline,array}

\usetikzlibrary{arrows,positioning,decorations.markings,decorations.pathmorphing,calc}
\definecolor{hyperref}{RGB}{026,028,087}

\def\gsim{ \lower .75ex \hbox{$\sim$} \llap{\raise .27ex \hbox{$>$}} }
\def\lsim{ \lower .75ex \hbox{$\sim$} \llap{\raise .27ex \hbox{$<$}} }
\def\be{\begin{equation}}
\def\ee{\end{equation}}
\def\bea{\begin{eqnarray}}
\def\eea{\end{eqnarray}}

\newcommand{\comment}[1]{}

\def\nn{\nonumber}

\definecolor{Gray}{gray}{0.9}
\definecolor{LightCyan}{rgb}{0.88,1,1}

\newcommand*{\mathcolor}{}
\def\mathcolor#1#{\mathcoloraux{#1}}
\newcommand*{\mathcoloraux}[3]{%
\protect\leavevmode
\begingroup
\color#1{#2}#3%
\endgroup
}
\newlength{\stheight}
\newcommand\textst[1][fu-grey]{
\ifmmode\setlength{\stheight}{+1.0ex}
\else\setlength{\stheight}{+0.5ex}
\fi
\bgroup\markoverwith{\textcolor{#1}{\rule[\the\stheight]{2pt}{1.0pt}}}\ULon
} %
\newcommand{\textins}[2][fu-grey]{
\ifmmode\mathcolor{#1}{#2}
\else\textcolor{#1}{#2}\@\,
\fi
}
\graphicspath{{./}}
\allowdisplaybreaks

\tikzstyle{vecArrow} = [thick, decoration={markings,mark=at position
1 with {\arrow[semithick]{open triangle 60}}},
double distance=1.4pt, shorten >= 5.5pt,
preaction = {decorate},
postaction = {draw,line width=1.4pt, white,shorten >= 4.5pt}]
\begin{document}

\hypersetup{pageanchor=false} 
\title{Hairy Black-holes in Shift-symmetric Theories}

\author[a,b]{Paolo Creminelli}
\author[c]{Nicol\'as Loayza}
\author[d,e]{Francesco Serra}
\author[d,e]{Enrico Trincherini}
\author[d,e]{Leonardo G. Trombetta}

\affiliation[a]{ICTP, International Centre for Theoretical Physics, Strada Costiera 11, 34151, Trieste, Italy}
\affiliation[b]{IFPU - Institute for Fundamental Physics of the Universe, Via Beirut 2, 34014, Trieste, Italy}
\affiliation[c]{IFIC, Universitat de Valencia - CSIC, E-46100, Burjassot, Spain}
\affiliation[d]{Scuola Normale Superiore, Piazza dei Cavalieri 7, 56126, Pisa, Italy}
\affiliation[e]{INFN - Sezione di Pisa, 56100 Pisa, Italy}

\emailAdd{creminel@ictp.it}
\emailAdd{nicolas.loayza@uv.es}
\emailAdd{francesco.serra@sns.it}
\emailAdd{enrico.trincherini@sns.it}
\emailAdd{leonardo.trombetta@sns.it}

\abstract{Scalar hair of black holes in theories with a shift symmetry are constrained by the no-hair theorem of Hui and Nicolis, assuming spherical symmetry, time-independence of the scalar field and asymptotic flatness. The most studied counterexample is a linear coupling of the scalar with the Gauss-Bonnet invariant. However, in this case the norm of the shift-symmetry current $J^2$ diverges at the horizon casting doubts on whether the solution is physically sound. We show that this is not an issue since $J^2$ is not a scalar quantity, since $J^\mu$ is not a diff-invariant current in the presence of Gauss-Bonnet. The same theory can be written in Horndeski form with a non-analytic function $G_5 \sim \log X$. In this case the shift-symmetry current is diff-invariant, but contains powers of $X$ in the denominator, so that its divergence at the horizon is again immaterial. We confirm that other hairy solutions in the presence of non-analytic Horndeski functions are pathological, featuring divergences of physical quantities as soon as one departs from time-independence and spherical symmetry. We generalise the no-hair theorem to Beyond Horndeski and DHOST theories, showing that the coupling with Gauss-Bonnet is necessary to have hair.}


\notoc
\maketitle
\newpage

\section{Introduction}

Do black holes have hair? Fifty years have passed since this question was first formulated but it is still able to fuel new ideas. 
One of the reasons behind its longevity is that both the theoretical and the experimental context surrounding it have changed dramatically in the last half a century. For instance, while the original emphasis was on characterizing the possible existence of additional parameters---in addition to the black hole mass, charge and angular momentum---that can be seen from far away, after the beginning of the era of gravitational wave astronomy, a more promising perspective has come out. Indeed, the presence of a non-trivial background at length scales of order of the light ring can modify the quasi normal mode spectrum and leave a detectable imprint in the black hole ringdown, which can therefore serve as a window on the dynamics of the gravitational sector. At least from this point of view, today there is no reason to prefer long over short hair. 

Anyway, in this paper we will readdress once again the aforementioned question, focusing on a somehow specific though, we think, significant situation. We will restrict to the case of {\it scalar} hair in {\it shift-symmetric} theories. Indeed if a scalar field is, with the exception of a cosmological constant, the most plausible ingredient to be added to General Relativity to explain e.g.~the accelerated expansion of the Universe, the presence of a shift symmetry represents the minimal choice to guarantee that such a field will be almost massless and hence relevant on cosmological scales. 

Within this specific setting a clear answer, for spherically symmetric and time-independent solutions going to a constant asymptotically\footnote{In this paper we do not consider the possibility of an asymptotic time-dependent solution $\phi = c \cdot t$.}, was obtained in a nice paper by Hui and Nicolis \cite{Hui:2012qt}. In their proof that such black holes have no hair, a crucial role is played by the covariantly conserved current $ J^{\mu}$. The shift invariance $\phi \to \phi + c$ actually implies that the scalar equation of motion can be written as the conservation of a current which depends on the field only through its derivatives. Because of the assumed spherical symmetry and static nature of the scalar and metric backgrounds, the only non-vanishing component can be $J^r$. As we will explain in detail at the beginning of the next section, for a black hole solution to be consistent, at the horizon every {\it physical} and {\it local} scalar quantity, and $J_\mu J^\mu$ in particular, must be finite. This implies that $J^r$ has to vanish at the horizon, since $g_{rr}$ diverges on that surface. Using now the conservation of the current, Ref.~\cite{Hui:2012qt} argues that $ J^r\equiv 0 $ everywhere.  
At this point, the authors noticed that if the dependence on $ \phi $ in the Lagrangian starts quadratically then the current will be proportional to $ \phi' $:
\begin{align} \label{Jr-lambertos}
J^r=\phi' F[\phi',g,g']\quad \text{with }F\text{ a regular function}\;,
\end{align}
where the absence of $\phi''$ and higher derivatives of the field is ensured if the theory has second-order equations of motion.
The presence of a kinetic term for the scalar field translates in the fact that $ F $ approaches a non-zero constant as $ \phi'\to0 $.  These two facts, along with the condition $ J^r\equiv0 $, imply that if $ F[\phi',g,g'] $ is a regular function around $ \phi'=0 $, then it must be $\phi'\equiv0 $ and therefore the hair vanishes.

Soon after the appearance of this theorem, however, it was realized \cite{Sotiriou:2013qea} that such a simple and compelling argument admits a subtle exception. Consider for instance the theory
\be
S = \frac{M_{\rm Pl}^2}{2} \int d^4 x \sqrt{-g}  \left(R -\frac12 (\partial\phi)^2 + \alpha \phi {\cal R}^2_{\rm GB} \right) \;.
\ee
The Gauss-Bonnet invariant, ${\cal R}^2_{\rm GB} \equiv R^{\mu\nu\rho\sigma}R_{\mu\nu\rho\sigma}-4R^{\mu\nu}R_{\mu\nu}+R^2$, is a total derivative and thus its coupling with the scalar preserves the shift symmetry $\phi \to \phi + c$. This term gives a $\phi$-independent contribution to the scalar equation of motion (and therefore to the current $J^r$), invalidating the assumption in (\ref{Jr-lambertos}). It acts as a source in the $J^r = 0$ equation, that does no longer allow for the trivial solution with a vanishing $\phi'$. While the presence of a linear scalar Gauss-Bonnet (sGB) coupling is indeed a sufficient condition to guarantee that black holes have hair, the actual solution found in \cite{Sotiriou:2013qea} seems puzzling. In this case not only $J^r$ contains a $\phi$-independent term, which is enough to circumvent the conclusion of the theorem, but also the norm of the current diverges at the horizon, as pointed out in \cite{Babichev:2017guv}.

A natural concern at this point is whether such a divergence, despite the regularity of the stress-energy tensor and of the resulting geometry at the horizon, is enough to conclude that solutions sourced by the Gauss-Bonnet coupling are not physical and therefore that the no-hair result in shift-symmetric theories is robust. A more optimistic perspective could instead be that the sGB example is just the first manifestation of a whole class of theories with hairy black holes, sourced by Lagrangian operators that give rise to $\phi$-independent contributions to $J^\mu$, among which there can be solutions with finite $J^2$.

A reason to consider the second possibility is the following. While the sGB coupling manifestly contains terms with higher derivatives on the metric, it gives rise to second-order equations of motion. This means that it has to belong to the large family of shift-symmetric scalar-tensor Lagrangian with this property, the so-called Horndeski theories \cite{Horndeski:1974wa} or generalized galileons \cite{Deffayet:2011gz}. Such an equivalence was pointed out in \cite{Kobayashi:2011nu} and it will be discussed in App.~\ref{app:H5-sGB}. From this point of view, the peculiarity of the sGB operator grows dim and in fact Ref. \cite{Babichev:2017guv} finds several other operators of the Horndeski-type that give contributions to $J^r$ that depend only on the metric in spherically symmetric and static backgrounds. The authors then conclude that there exist examples of black holes with hair and a finite norm of the current at the horizon. Despite this result, in a subsequent paper \cite{Saravani:2019xwx} it is claimed that all the hairy black hole solutions of this kind cannot be smoothly connected to Minkowski space-time, leaving those generated by the Gauss-Bonnet linear coupling as the only possibility.   

Given the somehow unsettled status of the original question in the literature, in this paper we will try to clarify if hairy black holes in shift-symmetric theories are {\it One, No One or One Hundred Thousand} (quoting Pirandello). The first step will be to show (Section \ref{sec2}) that the Gauss-Bonnet current is not covariant under diffeomorphism and, as a result, $J^2$ is not a scalar quantity. Its divergence at the horizon is therefore non-physical. The existence of an equivalent description of the sGB coupling in terms of a Horndeski operator, however, implies that there is a different form of the current which is instead covariant and still divergent. In spite of that, as we will discuss in Section \ref{sec3}, in this case the vector $J^\mu$ and its norm contain powers of $(\partial\phi)^2$ at the denominator. This non-locality for $X \to 0$ does not affect in any way the dynamics, but deprives the divergence of $J^2$ of any physical meaning. The conclusion is that the presence of Gauss-Bonnet actually represents a well-defined exception to the no-hair theorem. Notice that only black holes feature ``long" hair in these theories, i.e.~solutions $\phi \propto 1/r$, while compact objects without horizon like neutron stars do not \cite{Yagi:2015oca}.

After having discussed the sGB models, in Section \ref{sec4} we move on and examine the whole class of Horndeski shift-symmetric Lagrangians to identify if a similar behaviour is present in other cases as well. While, as already noticed in \cite{Babichev:2017guv}, for static and spherically symmetric solutions there are several operators that contribute to $J^\mu$ with a regular and scalar-independent term, as soon as the background solution is slightly deformed, every operator of this type manifests its non-local nature and becomes divergent in the limit of Minkowski spacetime. To further assess the robustness of the no-hair result, in Section \ref{sec5} we then extend the analysis to Lagrangians that still propagate $3$ degrees of freedom (the graviton plus a scalar) but nonetheless have equations of motion with higher-order derivatives, the so called degenerate higher-order scalar-tensor (DHOST) theories. These include Horndeski and Beyond Horndeski as particular cases.

In Section \ref{sec53} we briefly study the case of the most general shift-symmetric EFT. In this class a prototypical example, which shares many similarities with sGB, is given by $\phi R \tilde{R}$. Finally, conclusions are drawn in Section \ref{sec6}.

\section{The Scalar Gauss-Bonnet Operator}
\subsection{The Gauss-Bonnet current}\label{sec2}
We want to understand whether the divergence of $J^2$ at the horizon is a pathology of the sGB hairy solutions or not. Why should the divergence of a scalar quantity $\cal{O}$ be worrisome, even when the stress-energy tensor and the geometry are regular at the horizon? One reason is that a scalar quantity can be added to the Lagrangian of the system with an arbitrary coefficient in front ${\cal L} \supset \lambda \cal{O}$. In doing so the black hole solution will change and if $\cal{O}$ diverges at the horizon, this will happen no matter how small $\lambda$ is\footnote{We thank M.~Mirbabayi for illuminating discussions about this point.}. The solution cannot be trusted since it is extremely ``unstable" if one modifies the theory. The situation is already pathological in classical physics, but it is even more so when we consider Quantum Mechanics, since loop corrections will induce $\lambda \neq 0$ even if we start with $\lambda =0$. Another related way to see the pathology is that in general a particle will be coupled to the scalar $\cal{O}$. This means that one gets an effect on the dynamics of the particle (and on its stress-energy tensor) that diverges at the horizon. This suggests that the solution is unstable when matter is included in the picture. 

What we said holds for a general operator $\cal{O}$, but $J^2$ turns out to be quite special. Indeed, the full current contains a part $J^\mu_{\rm GB}$ associated with the Gauss-Bonnet term in the action, $\phi \mathcal{R}^2_{\rm GB}$. This current {\it is not covariant under diffeomorphisms} and therefore any scalar built with it is not invariant under diffs. Therefore one cannot add to the Lagrangian $\lambda J^2$ (or write a coupling with a particle) and the issue above does not arise. Simply stated, $J^2$ is not a scalar quantity: its value, and thus its divergence, depends on the coordinates we choose. The divergence of $J^2$ is immaterial, like the divergence of a component of the metric or of a Christoffel symbol. The current $J^\mu_{\rm GB}$ satisfies $\nabla_\mu J^\mu_{\rm GB} = \mathcal{R}^2_{\rm GB}$, but the form of the current is ambiguous and there is no privileged expression, even when a coordinate system is chosen \cite{Yale:2010jy}. 

This statement is analogous to what happens in a (non-abelian) gauge theory for the term ${\rm Tr }F_{\mu\nu}\tilde F^{\mu\nu}$. This object is notoriously a total derivative ${\rm Tr }F_{\mu\nu}\tilde F^{\mu\nu} = \partial_\mu G^\mu$, with $G_{\mu} = \epsilon_{\mu\nu\lambda\sigma} {\rm Tr } A_\nu (F_{\lambda\sigma}-\frac23 A_\lambda A_\sigma)$. Similarly to our case, the current $G^{\mu}$ is not gauge-invariant and $G^2$ is not a gauge-invariant scalar that can be added to the Lagrangian.

Let us make some examples of the forms the current $J^\mu_{\rm GB}$ can take for the Schwarzshild metric. In the presence of Killing vectors, there is a simple way to write $J^\mu_{\rm GB}$ in the coordinates in which an isometry simply acts as a shift of one coordinate \cite{Yale:2010jy}. Suppose this coordinate direction has label $ W $, then 
\begin{align}\label{pad}
J^\mu_{\rm GB} = 2 {P^{W\mu\nu}}_{\rho}\Gamma^{\rho}_{\nu W} \;,
\end{align}
where $ P^{\mu\nu\rho\sigma}=\partial \mathcal{R}^2_{\rm GB}/\partial R_{\mu\nu\rho\sigma}$ and $ \Gamma $ is the Christoffel symbol. This expression holds only in coordinates where the translation in $W$ is an isometry. For the case of Schwarzschild, one can use this expression in the standard coordinates $(t,r,\theta,\varphi)$ either using $W = t$ or $W = \varphi$. In the first case one gets a current that points only in the radial direction (we temporarily suppress the subscript GB)
\be
{J^{\mu}_{(t)}}=\left (0\,,\,-\frac{4 r_s^2}{r^5}\,,\,0\,,\,0\right )\;, \qquad {J_{(t)}}^2 =\frac{16 r_s^4}{r^9 (r-r_s)} \;.
\ee
Where $ r_s $ is the Schwarzschild radius. This is exactly the current discussed in the Introduction and indeed ${J_{(t)}}^2$ diverges at the horizon, $r=r_s$. On the other hand, with the choice $W = \varphi$ one gets
\be\label{phicurrent}
{J^{\mu}_{(\varphi)}}=\left (0\,,\,\frac{4 r_s(r-r_s)}{r^5}\,,\,-\frac{8 r_s \cot (\theta )}{r^5}\,,\,0\right ) \;, \qquad 
{J_{(\varphi)}}^2 =\frac{16 r_s^2 \left(-r_s+4 r \cot ^2(\theta )+r\right)}{r^9} \;.
\ee
The divergence of both these currents gives the Gauss-Bonnet invariant:
$\nabla_{\mu}J^{\mu}_{(t)}=\nabla_{\mu}J^{\mu}_{(\varphi)}=\mathcal{R}^2_{\rm GB}=12r_s^2/r^6$.  However ${J_{(\varphi)}}^2$ is finite at the horizon, while it diverges on the azimuthal axis. This example makes clear that $J_{\rm GB}^2$ is not a diff invariant quantity.

A general expression of the Gauss-Bonnet current, which does not assume isometries of the metric, can be given in terms of the spin connection. In the Appendix \ref{app:GBJs} we compute this current in the Schwarzschild spacetime and show that it is not covariant by writing it in different coordinate systems, in particular in Kruskal-Szekeres coordinates where the metric is regular at the horizon.

\subsection{Horndeski form of sGB}\label{sec3}

Since the sGB operator is such that the equations of motion are of second order and that there is symmetry under constant shift of the scalar field, it must be possible to express it in terms of the so-called shift-symmetric Horndeski Lagrangian. Indeed, the latter describes the most general shift-symmetric scalar-tensor theory with second-order equations, and is given by the sum of the following terms:
\begin{align}\label{HLag}
{\cal L}_{2}^H & = G_2(X)\,, \nn\\
{\cal L}_{3}^H & = G_{3}(X)[\Pi]\,, \nn \\
{\cal L}_{4}^H  & = G_{4}(X) R - 2 G_{4,X}(X)\left([\Pi]^{2}-[\Pi^2]\right), \nonumber \\
{\cal L}_{5}^H & = G_{5}(X)G_{\mu\nu}\Pi^{\mu\nu} +\frac{1}{3} G_{5,X}(X)\left([\Pi ]^{3}-3[\Pi][\Pi^2]+2[\Pi^3]\right),
\end{align}
where $X \equiv g^{\mu\nu} \partial_\mu \phi \partial_\nu \phi$, $\Pi_{\mu\nu} \equiv\nabla_{\mu}\nabla_{\nu}\phi $, $G_{\mu\nu}$ is the Einstein tensor and square brackets indicate the trace of an expression, e.g.~$ [\Pi]=\Box\phi$. 

It has been pointed out in Ref.~\cite{Kobayashi:2011nu} that the choice $G_5 = \log(X)$ gives indeed the same equation of motion as the linear sGB operator (without any field redefinition). In Appendix \ref{app:H5-sGB} we give some details about the proof of this equivalence. 
The benefit of this alternative way of writing the sGB operator is that now the Noether current $J_{H5}^\mu$ associated with the shift-symmetry is covariant. For this reason, contrarily to the previous case, the norm of this current $(J_{H5})^2$ is a true scalar and its divergence looks now problematic. For $G_5 \sim \log{|X|}$ one has
 \begin{eqnarray} \label{J5_0_squared}
 \left(J_{H5} \right)^2 &=& \frac{4}{X^4} \Biggl\{ - \frac{X}{18} \Bigl( [\Pi]^3 - 3 [\Pi][\Pi^2] + 2[\Pi^3] \Bigr)^2 \\
 &&\qquad+ \partial \phi \cdot \Biggl[ \Pi^6 -2[\Pi]\Pi^5 + [\Pi]^2 \Pi^4 - \frac{\Pi^3}{3}\left( [\Pi]^3 - 3 [\Pi][\Pi^2] +2[\Pi^3]  \right) \notag\\
 &&\qquad\qquad\quad + \frac{\Pi^2}{12} \left( [\Pi]^4 - 6 [\Pi]^2[\Pi^2] + 8 [\Pi][\Pi^3] - 3[\Pi^2]^2 \right) \notag\\
 &&\qquad\qquad\quad + \frac{\Pi}{6} \left( [\Pi^2] - [\Pi]^2 \right) \left( [\Pi]^3 - 3 [\Pi][\Pi^2] +2[\Pi^3]  \right) 
 \Biggr] \cdot \partial \phi \Biggr\} + {\cal O}(R_{\mu\nu\rho\sigma}) \,. \notag
\end{eqnarray}
Here we only wrote explicitly the terms that survive in flat space: the complete expression contains terms up to quadratic order in the curvature, indicated by ${\cal O}(R_{\mu\nu\rho\sigma})$.
It is easy to see that $(J_{H5})^2$ is a non-local operator, with powers of $X$ at denominator. As such it cannot be added to the action if one is interested in solutions for which $X \to 0$ somewhere. Therefore, its divergence is immaterial, in the same way one is not worried about $1/X$ going to infinity for a solution where the scalar is a constant. The above quantity is generally ill-defined as $X \to 0$. As discussed in Ref.~\cite{Babichev:2017guv}, $(J_{H5})^2$ diverges on the horizon of a hairy black hole. However this does not invalidate the solution, since the operator is non-local. 

Since we now understand that the operator $(J_{H5})^2$ is non-local and cannot be added to the Lagrangian, one may worry about the theory we started with, featuring $G_5 = \log(X)$. The appearance of powers of $X$ in denominators suggests that the theory is pathological in the limit $X \to 0$. However, this cannot be the case, since the theory is equivalent to the original sGB. Indeed, the non-locality of $G_5 = \log(X)$ is only apparent as we are going to explicitly show in Section \ref{sec4} and Appendix \ref{app:H5-sGB}.

\subsection{Boundedness of local scalar quantities}

Having established that the divergence of a nonlocal quantity does not invalidate the sGB solution, one may ask whether there can be instead a local scalar quantity that diverges. One can then fully trust the solution only if no local scalar operators blow up (outside the physical singularities). Here we will verify that this is indeed the case for the sGB solution. We will see that requiring the boundedness of scalar quantities forces a condition on the scalar field $\phi$, i.e. that all its radial partial derivatives $ \partial^n_r\phi$ have to be bounded everywhere, and in particular at the black hole horizon. 

While this result is evident far away from the black hole, it becomes less obvious at the horizon, where some coordinate systems display a non-physical singularity.
In order to overcome this complication, one can choose coordinates in which all the geometrical quantities are smooth at the horizon. This can be achieved for instance by means of Kruskal-Szekeres-like coordinates or through locally inertial coordinates, where the metric is set to Minkowski in a specific point (e.g.~$ g_{\mu\nu}(p)=\eta_{\mu\nu} $ in a point $ p $ at the horizon). Choosing these last coordinates, the Christoffel symbols will vanish at the chosen point but will have non zero derivatives: these will describe (up to a Lorentz boost) the finite tidal forces experienced by a free-falling observer that is crossing the horizon. 

Being interested in spherically symmetric and static solutions, it is enough to compute scalar quantities in a single point of the horizon. Moreover, since the geometry of a black hole is non-singular at the horizon, one can simply consider quantities that depend on the scalar field, for instance having the form $ (\nabla^n\phi)^2 $. For the same reason, in these quantities the terms displaying the most severe divergence when derivatives of $ \phi $ are not well behaved will be those involving only partial derivatives\footnote{Even though divergent boost factors might make the derivatives of the Christoffel symbols diverge, these would still be subleading with respect to partial derivatives of the scalar field, which would get the same boost-enhancement.}.

Writing the metric in Schwarzschild-like coordinates as
\begin{align}
ds^2=-f(r)dt^2+\frac{dr^2}{f(r)}+\rho^2(r)(d\theta^2+\sin^2\!\!\theta\, d\varphi^2)\,,
\end{align}
with $ f=0 $ for $ r=r_s $, we can define a locally inertial frame in a point $ p $ using coordinates $ (\hat{t},\hat{r},\hat{\theta},\hat{\varphi}) $ having origin in $ p $ and such that in $p$:
\begin{align}
d\hat{t}=\sqrt{f}dt\;,\; d\hat{r}=\frac{1}{\sqrt{f}}dr\;,\; d\hat{\theta}=\rho\,d\theta\;,\; d\hat{\varphi}=\rho\,\sin\!\theta\,d\varphi\;.
\end{align}
The Jacobian of this transformation will be diagonal in $ p $. For this reason we understand that in the leading term of $ (\nabla^n\phi)^2 $ in $ p $ only $ \hat{r} $ partial derivatives will appear, each corresponding to a weighted $ r $ partial derivative: $\partial_{\hat{r}}=\sqrt{f}\partial_r$ . In conclusion in the chosen point we have
\begin{align}\label{nablansq}
\left.(\nabla^{n}\phi)^2\right.\sim \left.(\partial_{\hat{r}}^n\phi)^2\right.+\dots\sim \left.f^{n}(\partial_r^n\phi)^2\right.+\left.f^{n-k-p}\partial_r^{n-p}\phi\partial_r^{n-k}\phi\right.+\dots\;,
\end{align}
where the second term on rhs (with $ k+p< n-1 $) indicates schematically a series of contributions having the same magnitude of the first one and the dots indicate smaller terms where derivatives hit the Christoffel symbols.

Writing $\phi\sim f^{\gamma} $ when $r \sim r_s$, it becomes clear that if $ \gamma $ is not a positive integer (or zero) there will be large enough values of $ n $ such that the terms in Eq.~\eqref{nablansq} will diverge at the horizon\footnote{For some special non-integer values of $ \gamma $ there will be a single integer $ n_{\gamma} $ for which the leading contributions in \eqref{nablansq} add up to zero, but this does not change our conclusion, since infinitely many other scalars will diverge.}, making the whole scalar $ (\nabla^n\phi)^2 $ diverge as $ f^{2\gamma-n} $. For this reason we see that all the scalar quantities built using the metric and the scalar field will be bounded when computed on a sGB hairy background if the hair has $ \partial_r\phi $ and its higher radial derivatives bounded at the horizon. This condition is satisfied by the perturbative solution of \cite{Sotiriou:2014pfa}.

\section{Additional hair in Horndeski?}\label{sec4}
In the Horndeski form, the sGB theory violates the assumptions of the no-hair theorem of \cite{Hui:2012qt} since $G_5$ in non-analytic for $X \to 0$. Indeed the current does not start linearly in $ \phi'$  and Eq.~\eqref{Jr-lambertos} does not hold. A natural question is therefore whether one can find additional hairy solutions (under the same symmetry assumptions stated in the Introduction) when the other Horndeski functions are non-analytic for $X \to 0$.  Examples of such theories have already been considered in  \cite{Babichev:2017guv}, where some particular cases were studied in which the radial component of the current contains a $ \phi' $-independent term (i.e.~a term in $ F[\phi',g] $ proportional to $ 1/\phi' $), in a static and spherically symmetric setting. In this section (see also \cite{Saravani:2019xwx}) we study this possibility and we conclude that all these additional examples are pathological. Here we stick to theories with second-order equations of motion, while more general cases will be discussed in the next section. 

Writing the spherically symmetric, static metric as
\begin{equation}
 ds^2 = -h(r) dt^2 + \frac{dr^2}{f(r)} + r^2 d\Omega^2\;,
\end{equation}
the radial component of the shift-symmetry Noether current for generic Horndeski Lagrangian \eqref{HLag} takes the form \cite{Babichev:2017guv}:
\begin{eqnarray}\label{Horndeski-J}
J_{H}^r &=& 2 f \phi' G_{2X} + f \frac{r h' + 4h}{rh} X G_{3X} - 4 f \phi' \frac{fh - h + rf h'}{r^2 h} G_{4X} - 8 f^2 \phi' \frac{h + rh'}{r^2 h} X G_{4XX}  \notag \\ &&
 - f h' \frac{1-3f}{r^2h} X G_{5X} + 2 \frac{h' f^2}{r^2 h} X^2 G_{5XX}\;.
\end{eqnarray}
Therefore, we see that it is possible to have contributions independent of $ \phi' $ when the functions $ G_i $ behave at small $X$ as
\begin{align}\label{specialGs} G_2(X) \sim \sqrt{|X|}\,,\;\;G_3(X) \sim \log|X|\,,\;\;G_4(X) \sim \sqrt{|X|}\,,\;\;G_5(X) \sim \log{|X|}\,,
\end{align} 
where the last choice gives the sGB operator. Keep in mind that the function $ G_4 $ will always include a leading constant term that drops out of the current in Eq.~$\eqref{Horndeski-J}$ and corresponds to the Einstein-Hilbert part of the Lagrangian.
The non-analytic behaviour for $X \to 0$ is worrisome when one wants to study the Lorentz-invariant vacuum $X =0$ or approaching it as it happens going far away from a localised black-hole solution. In the following we are going to show that these theories are indeed pathological, with the only exception of sGB.

\subsection{Troubles with a Lorentz-invariant solution}
We want to study Lorentz-invariant solutions of the theories with the non-analytic behaviours of Eq.~\eqref{specialGs}. We are going to show that, with the exception of sGB, these solutions are pathological since the equations of motion are not continuous in this limit, i.e.~the result depends on how the flat, Lorentz invariant solution is approached. Let us take the metric to be Minkowski from the beginning $g=\eta$ (this defines a particular direction in which we approach the solutions we are interested in).

The equations of motion for generic Horndeski functions $G_i$ read:
\begin{eqnarray}\label{eoms-H-flat}
\nabla_{\mu} J_{H2}^{\mu}\Big|_{g=\eta} &=& 2 G_{2,XX}\, \Pi^{\mu\nu} \, \partial_\mu \phi\,\partial_\nu \phi  + G_{2,X} [\Pi] \;, \nonumber\\
\nabla_{\mu} J_{H3}^{\mu}\Big|_{g=\eta} &=& 4 G_{3,XX} \left[ \Pi^{\mu\nu}\, [\Pi] - (\Pi^2)^{\mu\nu} \right] \partial_\mu \phi \, \partial_\nu \phi + 2 G_{3,X} \left( [\Pi]^2 - [\Pi^2] \right)\;, \nonumber\\
 \nabla_{\mu} J_{H4}^{\mu}\Big|_{g=\eta} &=& 8 G_{4,XXX} \left[ (\Pi^3)^{\mu\nu} - (\Pi^2)^{\mu\nu}\, [\Pi] + \frac{1}{2} \Pi^{\mu\nu} \left( [\Pi]^2 - [\Pi^2] \right) \right] \partial_\mu \phi \,\partial_\nu \phi \notag \nonumber\\
 &&+ 2 G_{4,XX} \left( [\Pi]^3 - 3 [\Pi^2][\Pi] + 2[\Pi^3] \right) \;, \nonumber\\
 \nabla_{\mu} J_{H5}^{\mu}\Big|_{g=\eta} &=& 4 G_{5,XXX} \Bigl[ (\Pi^4)^{\mu\nu} - (\Pi^3)^{\mu\nu}\, [\Pi] + \frac{1}{2} (\Pi^2)^{\mu\nu} \left( [\Pi]^2 - [\Pi^2] \right) \notag \nonumber\\
 &&\qquad \qquad \qquad \qquad \quad\,\,\,- \frac{1}{6} \Pi^{\mu\nu} \left( [\Pi]^3 - 3[\Pi][\Pi^2] + 2 [\Pi^3] \right) \Bigr] \partial_\mu \phi \,\partial_\nu \phi \notag \nonumber\\
 &&-\frac{1}{3} G_{5,XX} \Bigl( [\Pi]^4 - 6[\Pi^2][\Pi]^2 + 3[\Pi^2]^2 + 8[\Pi^3][\Pi] -6[\Pi^4] \Bigr)\;.
\end{eqnarray}
When the functions $ G_i,\;i=2,3,4,5 $ behave as in Eq.~\eqref{specialGs}, the above equations take the form
\begin{equation}
 \nabla_{\mu} J_{Hi}^{\mu}\Big|_{g=\eta} \sim \frac{1}{X^{(i+1)/2}} \left( A_{i}^{\mu\nu} \, \partial_{\mu} \phi \, \partial_{\nu} \phi + [A_{i}] \, c_i X \right)\;,
\end{equation}
where $A_{i}^{\mu\nu}$ are tensors built out of the $(i-1)$-th power of $\Pi^{\mu\nu}$, and $c_i$ are numerical coefficients that can be easily determined by inspection:
\begin{equation}
c_i = - \frac{1}{i-1}\;.
\end{equation}
Notice that all of the above equations of motion (in flat-space) are finite for time-independent and spherically symmetric backgrounds, as it can be confirmed by taking the divergence of \eqref{Horndeski-J} when the $G_i$'s are given by \eqref{specialGs} and then taking the Minkowski limit $f,h \to 1$. This is why no apparent problem arises when looking for hairy black-hole solutions. 

However, if a Lorentz invariant solution were to exist in $d$ dimensions, then one would have $A_{i}^{\mu\nu} = \eta^{\mu\nu} [A_i]/d$ (this can be seen as an additional assumption about the direction in which the limit is approached), and therefore the equation would simplify to
\begin{equation} \label{eom-H-lorentz} 
\nabla_{\mu} J_{Hi}^{\mu}\Big|_{g=\eta} \propto \left( \frac{1}{d} + c_i \right) \frac{[A_{i}]}{X^{(i-1)/2}} = P_i(d) \left( \frac{1}{d} + c_i \right) \left(\frac{[\Pi]}{X^{1/2}} \right)^{i-1}.
\end{equation}
In the last expression we used $\Pi^{\mu\nu} = \eta^{\mu\nu} [\Pi]/d$, with the prefactor given by $P_i(d) = \prod_{p=0}^{(i-2)} (d-p)$. Since $[\Pi]/X^{1/2} \sim \partial^2 \phi/\partial \phi$, one has the same number of fields at numerator and denominator, so that the Lorentz-invariant limit is ambiguous. Consider for instance $\phi = A \, x_{\mu} x^\mu + b_\mu x^\mu + c$, for which $\Pi_{\mu\nu} = 2A\, \eta_{\mu\nu}$. The trivial Lorentz-invariant and translationally invariant configuration $\phi = const$ is reached when $A \to 0$ and $b_\mu \to 0$. Expressions like $[\Pi]/X^{1/2}$ depend on the order of these limits.

However, for each operator there is a critical dimension for which Eq.~\eqref{eom-H-lorentz} identically vanishes, namely
\begin{equation}\label{dCond}
 d_i = (i-1)\;.
\end{equation}
This means that except for a single value of $i$, all of the other cases\footnote{Here we mean those that are not automatically trivial. As it is well known \cite{Nicolis:2008in}, for a given dimension $d$, the Galileon-like structures present in Horndeski theories with $i > d-2$ are indeed trivial. In our setup this can be seen in Eq. \eqref{eom-H-lorentz} from the fact that $P_{(d+2)}(d) = 0$.} are incompatible with a Lorentz invariant solution.
This analysis is enough to conclude that in $d=4$ all of the cases in \eqref{specialGs} are not compatible with a Lorentz-invariant solution with the exception of sGB (in Appendix \ref{app:H5-sGB} we are going to also study the $d=2$ case with $G_3 \sim \log{|X|}$, which corresponds to a coupling $\phi ^{(2)}\!R$).

\subsection{Troubles with perturbations}

A similar situation arises when considering arbitrary perturbations around an $ X=0 $ background. For simplicity we will consider a Lorentz-invariant one. Indeed, consider a scalar quantity built with the scalar fields's first and second derivatives, $\mathcal{O}(\partial \phi, \Pi)$. Expanding in linear perturbations, $\phi = \phi_0 + \delta \phi$, it takes the form
\begin{equation}
 \delta \mathcal{O} = B^\mu \, \partial_\mu \delta\phi + C^{\mu\nu} \, \partial_\mu \partial_\nu \delta\phi\;,
\end{equation}
where $B^\mu$ and $C^{\mu\nu}$ depend on background quantities only, and for a Lorentz invariant background will satisfy
\begin{equation}
 B^\mu = 0 \qquad ; \qquad C^{\mu\nu} \propto \eta^{\mu\nu}.
\end{equation}
Therefore, it is enough to only track the perturbations with two derivatives acting on $ \delta\phi $ . For example, linear perturbations of the equations of motion \eqref{eoms-H-flat} are
\begin{equation} \label{eom-H-lorentz-pert}
 \delta \left( \nabla_{\mu} J_{Hi}^{\mu}\Big|_{g=\eta} \right) = \mathcal{Z}_{i} \,  \square \delta\phi\;,
\end{equation}
with
\begin{equation}\label{Zis}
 \mathcal{Z}_{i} \propto   \frac{P_{i+1}(d)}{\sqrt{X}} \left(\frac{[\Pi]}{X^{1/2}} \right)^{i-2}\;.      
\end{equation}
Again, we observe a problem in the limit $\phi \to const$ for the cases in \eqref{specialGs} which is now even worse than for the background equations \eqref{eom-H-lorentz}, since here there is an extra power of the field's first derivatives in the denominator. Also, similarly to what happened for the background equations discussed above, in $d=4$ dimensions we see that the choice $G_5(X)\sim\log|X|$, i.e.~sGB, is safe because the prefactor $P_{6}(4)$ vanishes (in a similar way in $d=2$ we have an analogous result for the cubic Horndeski $P_4(2) =0$). 
Of course $G_5(X)\sim\log|X|$ would continue to avoid problems, even going to higher order in perturbations and on more general backgrounds. Indeed, as we discussed, this case does not feature any true non-locality being equivalent to the sGB theory (see Appendix \ref{app:H5-sGB}).

It is important to point out that, besides the cases \eqref{specialGs}, many other choices of the $ G_i $ can produce hairy solutions, as long as $ J^r $ contains terms less than linear in $ \phi' $, so that Eq.~\eqref{Jr-lambertos} does not hold. One such example is $ G_3(X)\sim X^{1/4} $, which produces a term proportional to $ \sqrt{\phi'} $. Even if in this case $ \phi'=0 $ solves $ J^r=0 $, from the explicit expression of $J^r$, Eq.~\eqref{Horndeski-J} one finds also a non-zero solution: \begin{align}
\phi'\propto\frac{f^{1/2}(rh'+4h)^2}{r^2 h^2}\quad \sim\frac{1}{r^2}\quad \text{ as }\;r\to\infty \;.
\end{align}
However, the same analysis carried out above for the cases \eqref{specialGs} shows that also this case is pathological. The analogue of Eq.~\eqref{Zis} now reads
\be
\tilde{\mathcal{Z}}_{3} = \frac{4}{d} (d - 1) [\Pi] \left( \frac{2}{d} X G_{3XX} + G_{3X} \right) \sim \frac{[\Pi]}{X^{3/4}}
\ee
and again the Lorentz-invariant limit is not well-defined. These pathologies will arise for any non-analytic function at a certain order in perturbations. For instance even an apparently innocuous term $ X^{n+1/2} $ will get corrections of the form $ \sim X^{n-k+1/2}(\delta X)^k $ when we consider deformations $ X\mapsto X+ \delta X $ of the background solution.
These terms will diverge as soon as $ k>n $, making impossible to compute corrections whenever $ X= 0 $, both on the Lorentz invariant vacuum and on hairy solutions.

In conclusion, by dropping the assumption of Eq.~\eqref{Jr-lambertos} one gets healthy hairy solutions only in the case of sGB.
The physical validity of the hairy black-hole solutions in theories of the form of Eq.~\eqref{specialGs} was studied in \cite{Saravani:2019xwx}, reaching a similar result. However, the arguments of \cite{Saravani:2019xwx} are not completely conclusive in our view. The authors point out that if one sets $\phi = const$, with a spherically symmetric and static metric and takes the limit of Minkowski spacetime, $J^r$ goes to zero only in the sGB case. This can be easily checked in the explicit expression of $J^r$ of Eq.~\eqref{Horndeski-J}.
However, in the case $G_4(X) \sim \sqrt{|X|}$ one gets $J^r \propto r^{-2}$ and this does not contribute to the equation of motion $\nabla_\mu J^\mu =0$. (Notice that a static solution is effectively 3-dimensional, so that, following the argument of Eq.~\eqref{dCond}, it is not surprising that the $G_4 \sim \sqrt{|X|}$ case is healthy for a static solution.) Actually, as we discussed at length in the previous sections, $J^r = 0$ is not a necessary requirement when $J^2$ is a non-local operator, as it is the case for all the choices in Eq.~\eqref{specialGs}, including sGB. As our analysis shows, one needs to go beyond static solutions to pinpoint the pathology. This also allows to exclude cases like $ G_3(X)\sim X^{1/4} $ discussed above, which were not covered by the arguments of \cite{Saravani:2019xwx} since the current vanishes once $\phi' =0$ is taken.

\section{Theories with higher-order equations of motion}\label{sec5}

So far we focussed on shift-symmetric theories with second-order equations of motion (Horndeski). However, the requirement that the field equations are of second order, which ensures there are no ghost degrees of freedom, can be relaxed. Indeed, even scalar-tensor theories leading to higher-order equations of motion can, in some cases, propagate only gravity plus a single extra scalar degree of freedom. For instance this happens when the following (shift-symmetric) Beyond Horndeski Lagrangian \cite{Gleyzes:2014dya} is added to the Horndeski one \eqref{HLag}: 
\begin{align}\label{BHLag}
{\cal L}_{4}^{BH}  & = - F_4(X) {\epsilon^{\mu\nu\rho}}_\sigma\epsilon^{\mu'\nu'\rho'\sigma}\partial_{\mu}\phi\, \partial_{\mu'}\phi \, \Pi_{\nu\nu'}\Pi_{\rho\rho'}  \, ,\notag \\
{\cal L}_{5}^{BH} & =  - F_5(X) \epsilon^{\mu\nu\rho\sigma}\epsilon^{\mu'\nu'\rho'\sigma'}\partial_{\mu}\phi\, \partial_{\mu'} \phi\, \Pi_{\nu\nu'}\Pi_{\rho\rho'} \Pi_{\sigma\sigma'}  \, ,
\end{align}
provided this degeneracy condition is satisfied  \cite{Gleyzes:2014qga}:
\begin{equation}\label{deg-cond-BH}
X G_{5X} F_4 = 3F_5 \left[ G_4 - 2 X G_{4X}\right]\;.
\end{equation}
There is an even larger set of such theories known as DHOST \cite{Langlois:2015cwa, Crisostomi:2016tcp}, which includes both Horndeski and Beyond Horndeski as special cases. 
In the following we are going to extend the study of black-hole hair to this more general setup, always with the same symmetry assumptions made in the Introduction. Notice that the application of the no-hair theorem is now not obvious, since now one expects the radial component of the current to also depend on $\phi''(r)$, violating Eq.~\eqref{Jr-lambertos}. 

\subsection{No-hair theorem for DHOST}
Let us start with the class of DHOST theories that can be obtained via invertible conformal and disformal trasformations that depend on the scalar field:
\begin{equation}
\bar{g}_{\mu\nu} = \Omega(X) \, g_{\mu\nu} + \Gamma(X) \, \partial_\mu \phi \,\partial_\nu \phi \label{disf-transf} \;.
\end{equation}
The dependence of $\Omega$ and $\Gamma$ on $X$ only (and not on $\phi$) ensures that the shift-symmetry is preserved. (Notice that the scalar field is not changed in the transformation.) The kinetic term transforms as
\begin{equation} \label{barXofX}
\bar{X} = \frac{X}{\Omega + X \Gamma}\;.
\end{equation}
This relation with the Horndeski theories is a way to understand why these DHOST theories must propagate only gravity plus a single extra scalar degree of freedom. In particular, from Quartic and Quintic Horndeski one generates  \cite{Achour:2016rkg,BenAchour:2016fzp}
\begin{eqnarray}
\bar{\mathcal{L}}_4^H[\bar{G}_4] &=& \mathcal{L}_4^H[G_4] + \mathcal{L}_4^{BH}[F_4] + \sum_i \, \alpha_i L_i^{(2)}, \label{horndeski-class-4} \\
\bar{\mathcal{L}}_5^H[\bar{G}_5] &=& \mathcal{L}_5^H[G_5] + \mathcal{L}_5^{BH}[F_5] + \sum_j \, b_j L_j^{(3)}, \label{horndeski-class-5}
\end{eqnarray} 
where the $\alpha_i$'s and $b_j$'s are functions which parametrize the part of the DHOST Lagrangian which is neither Horndeski nor Beyond Horndeski\footnote{These functions are not all independent, but satisfy relations in order to ensure the degeneracy conditions analogous to \eqref{deg-cond-BH}.}, and $L_i^{(2)}$ and $L_i^{(3)}$ are terms quadratic and cubic in second derivatives of the scalar field respectively.

Let us consider first a theory with only Horndeski and Beyond Horndeski: it is generated by a purely disformal transformation, i.e.~$\Omega(X) = 1$ and $\Gamma(X) \neq 0$. Since the equations of motion are of higher order, one would expect $J^r$ to contain more derivatives
with respect to the form of Eq.~\eqref{Jr-lambertos}. However, this does not happen, as a consequence of the high degree of symmetry, and $\phi''$ does not appear in $J^r$ \cite{Babichev:2017guv}:
\begin{eqnarray} \label{Jr-spherical-BH}
J_{BH}^r &=& 4f^2 \phi' \frac{h+rh'}{r^2 h} X (2 F_4 + X F_{4X}) + 3 \frac{f^2 h'}{r h} X^2 (5F_5 + 2X F_{5X})\;.
\end{eqnarray}
Therefore the no-hair theorem applies without any changes. We will discuss below new exceptions in the same vein of Eq.~\eqref{specialGs}.

More generally, turning on the conformal part of the transformation, i.e.~$\Omega_{,X} \neq 0$, allows to span this full DHOST class. In this case the current will contain higher derivatives of the scalar field: these arise from derivatives of the metric, once one uses the transformation of Eq.~\eqref{disf-transf}. Therefore extra derivatives come from the derivatives of $ \Omega(X) $ and $ \Gamma(X) \partial_\mu\phi \partial_\nu\phi$. Assuming that the functions $\Omega$ and $\Gamma$ are regular for $X \to 0$ ($\Gamma$ must start as a constant and $\Omega$ as a non-zero constant) extra derivatives of $\phi$ will always appear alongside extra powers of $\phi$. Therefore the current, instead of being of the form of Eq.~\eqref{Jr-lambertos} is of the form\footnote{There are no terms with three or more derivatives of $\phi$ in the current, because they would give terms with four or more derivatives in the equations of motions. However the transformation \eqref{disf-transf} adds at most one derivative: starting with second-order equations, one ends up with at most three derivatives.}
\be
{J}^r=\phi' F[\phi',\phi'',g'] \;.
\ee
Now we are in the position of extending the theorem to this case. Since in any EFT derivatives must be bounded, in the limit $\phi' \to 0$ we also have $\phi'' \to 0$. In this limit the function $F$ must go to a constant as in the original case, since the new terms in the current are at least quadratic in $\phi$. Therefore the logic of \cite{Hui:2012qt} still applies: since $J^r =0$ (with the caveat of Gauss-Bonnet that we discussed at length) and $\phi' = 0$ asymptotically, it must remain so everywhere because for small values of the field the current is simply proportional to $\phi'$ so that this cannot move away from zero. In conclusion, the no-hair theorem is extended to DHOST theories which are connected to a healthy Horndeski theory (as defined in the previous Section) by means of a transformation with $\Omega(X)$ and $\Gamma(X)$ regular around $X=0$.

\subsection{The fate of sGB}

Another way to see that the theorem still holds is to look at how black-hole solutions are transformed. Since the scalar field is not changed by the transformation, hair can neither be generated nor removed (grown nor cut) by these transformations. Moreover, the asymptotics of the solutions are preserved and their symmetries as well. Indeed, far away from the black hole the transformation \eqref{disf-transf} becomes trivial ($\partial_\mu \phi \to 0$),
\begin{equation} \label{trivial-transf}
\bar{g}_{\mu\nu} = \Omega(0) \, g_{\mu\nu} \qquad (r\to \infty),
\end{equation}
where of course $\Omega(0) > 0$. This is a constant overall rescaling of the metric: spacetime is still asymptotically flat. 
Therefore the only DHOST theories with hair are the ones obtained via \eqref{disf-transf} starting from a sGB Horndeski theory, since this is the only Horndeski theory with hair.  (Here we are not considering the possibility that a black-hole solution is mapped into a solution with a naked singularity, as discussed in \cite{BenAchour:2019fdf}.)

The new terms generated by such transformation from both Quartic and Quintic Horndeski, Eqs.~\eqref{horndeski-class-4} and \eqref{horndeski-class-5}, are given by
\begin{eqnarray} 
G_{4,X} &=& \bar{G}_{4,\bar{X}} \sqrt{\Omega} (\Omega + X \Gamma)^{1/2}\;, \label{G4-transf} \\
F_4 &=& - \bar{G}_{4} \frac{(\Gamma \Omega_X + \Omega \Gamma_X)}{\sqrt{\Omega}(\Omega + X \Gamma)^{1/2}} + 2 \bar{G}_{4,\bar{X}} \frac{\sqrt{\Omega} (X \Gamma_X - \Omega_X)}{(\Omega + X \Gamma)^{3/2}}\;, \label{F4-transf}\\
\alpha_5 &=& - \bar{G}_{4} \frac{2 \Omega_X (\Gamma \Omega_X + 2\Omega \Gamma_X)}{\Omega^{3/2}(\Omega + X \Gamma)^{1/2}} + 4 \bar{G}_{4,\bar{X}} \frac{ \Omega_X (2 X \Gamma_X - \Omega_X)}{\sqrt{\Omega} (\Omega + X \Gamma)^{3/2}}\;, \label{alpha5-transf}
\end{eqnarray}
for the Quartic part, while for the Quintic part
\begin{eqnarray}
G_{5,X} &=& \bar{G}_{5,\bar{X}} \frac{\sqrt{\Omega} \left[ \Omega - X ( \Omega_X + X \Gamma_X) \right]}{(\Omega + X \Gamma)^{5/2}}\;,  \label{G5-transf} \\
F_5 &=& -2 \bar{G}_{5,\bar{X}}\frac{\sqrt{\Omega} ( \Omega_X + X \Gamma_X)}{3(\Omega + X \Gamma)^{5/2}}\;, \label{F5-transf} \\
b_4 &=& \bar{G}_{5,\bar{X}} \frac{\sqrt{\Omega} \, \Omega_X}{3(\Omega + X \Gamma)^{5/2}}\;. \label{beta4-transf}
\end{eqnarray}
Due to the degeneracy conditions (see Refs. \cite{Achour:2016rkg,BenAchour:2016fzp}), the remaining $\alpha_i$ and $b_j$ are determined by the ones shown, and therefore contain no new information. 
Starting with a Horndeski theory with hair, i.e.~with $\bar{G}_5 = \log(\bar{X})$ (sGB) one wants to know whether it is possible to end up in a DHOST theory without the sGB term (and with all functions regular for $X \to 0$). From Eq. \eqref{G5-transf} it would seem that there is a possible choice of $\Omega$ and $\Gamma$ in the transformation such that $G_5$ is regular in $ X=0 $, namely
\begin{equation}
\left[ \Omega - X(\Omega_{,X} + X \Gamma_{,X}) \right] \to 0\;,
\end{equation}
at least linearly in $X$. However, as discussed in Ref.~\cite{Achour:2016rkg}, when the above combination vanishes the transformation admits a null eigenvector, i.e. it is not invertible and thus pathological. 

Ref.~\cite{Babichev:2017guv} studied Beyond-Horndeski theories which could be exceptions to the no-hair theorem, along the lines of \eqref{specialGs}. These exceptions involve special choices of the Beyond Horndeski functions,
\begin{align}\label{specialchoice} F_4(X) \sim |X|^{-3/2}\,,\;\;F_5(X)\sim |X|^{-2}\,.
\end{align} 
The transformation laws \eqref{F4-transf} and \eqref{F5-transf} show however that these are not reachable with regular transformations, neither starting from regular Horndeski functions, nor allowing for sGB. Indeed, in the latter case one would need to allow for $\Gamma \sim X^{-1}$ in order to generate $F_5(X) = |X|^{-2}$ from $\bar{G}_5 = \log(\bar{X})$. It is straightforward to check that, although such transformation is safe in a static and spherically symmetric background, it is ill defined for a general configuration.

We conclude then that it is not possible to remove the sGB operator with a regular and invertible transformation of the form \eqref{disf-transf}. 
Therefore, the DHOST theories that we studied can be separated in two (invariant) subclasses, those with the sGB operator and therefore with hairy black holes and those without. In other words, for a given DHOST theory connected to Horndeski, in order to determine whether it can support healthy hairy black hole solutions or not, one only needs to check if Eqs. \eqref{G4-transf} to \eqref{beta4-transf} can be satisfied with $\bar{G}_5(\bar{X}) \sim \log(\bar{X})$ for small $\bar{X}$.

 \subsection{Other DHOST theories and beyond} \label{sec53}
Besides the theories discussed in the previous sections, other DHOST classes can be defined imposing different degeneracy conditions on the higher-derivative operators added to the Lagrangian \cite{Achour:2016rkg,BenAchour:2016fzp}. This procedure outlines various DHOST classes featuring operators either quadratic or cubic in second derivatives of the scalar field.  As discussed in \cite{Achour:2016rkg,BenAchour:2016fzp}, further requirements might be imposed in order to select the theories which can be interpreted as a modification of General Relativity through the presence of an additional scalar degree of freedom. 

In particular only some DHOST theories admit a ghost-free decoupling limit of the metric in flat spacetime. In addition to this, if one wishes to include operators from a cubic DHOST class, the degeneracy conditions required by these must be compatible with those of the quadratic DHOST theories which are necessary in order to include an Einstein-Hilbert term in the Lagrangian. As shown in \cite{BenAchour:2016fzp}, these two requirements narrow down the interesting classes to only two possibilities\footnote{Classes ${}^{2} \text{N-I}+{}^{3} \text{M-I}$ and ${}^{2} \text{N-I}+{}^{3} \text{N-I}$, as defined in Ref. \cite{BenAchour:2016fzp}.}. One of these is precisely the class studied in the previous sections, generated by conformal plus disformal invertible transformations of Horndeski theories\footnote{Non-invertible transformations will land either outside this class, or in a theory involving non-regular functions.}. The other class involves more complicated constraints and cannot be characterised as easily. In Appendix \ref{app:M3} we show that although this class accommodates both quadratic and cubic DHOST, it contains only theories that do not allow for an Einstein-Hilbert term and are therefore unsuitable to describe a modification of General Relativity.

One can consider an even more general situation. Imposing either second order or degenerate equations of motion is motivated if at least one higher derivative (HD) operator becomes large on the solutions one is interested in. On the other hand, if HD operators can always be treated perturbatively, as it typically happens in more conventional EFTs, then such a requirement is no longer necessary and arbitrary HD operators can be considered. (Notice that this possibility is not that different from the case of sGB discussed so far: even if the sGB gives second-order equations of motion, these equations may be pathological, featuring ghost or gradient instabilities, when the sGB is as important as the scalar kinetic term \cite{Ripley:2019hxt}.)

Interestingly the theorem of \cite{Hui:2012qt} can be extended to this very generic setting, as long as one considers energy scales below that at which the ghost degrees of freedom appear, i.e.~in the regime of validity of the EFT. In a spherically symmetric and static spacetime the current will take the form:
\begin{align}\label{genericcurr}
J^r=\phi' F_1+\phi'' F_2+...+\phi^{(n)}F_n\;,
\end{align}
where the functions $ F_i $ are assumed to be regular as $ \phi' $ and its derivatives approach zero. Sufficiently far away, and within the regime of validity of the EFT, the leading term will be the first one in Eq. \eqref{genericcurr}, so that following Ref. \cite{Hui:2012qt}, $J^r = 0$ implies $\phi = const$.

One can also find exceptions to this extension of the theorem, similarly to the case of sGB, where the current contains $\phi$-independent contributions. Among the various possible operators of this kind, the simplest example is given by $ \phi R\tilde{R} $, i.e. a linear coupling between the scalar field and the Chern-Simons topological density (see for example \cite{Okounkova:2017yby})
\begin{align}
\int d^4x\,\sqrt{-g}\, R\tilde{R}&=\int d^4x\,\sqrt{-g}\nabla_{\mu}K^{\mu}\\
K^{\mu}&=2\frac{\epsilon^{\mu\alpha\beta\gamma}}{\sqrt{-g}}\Gamma_{\alpha\sigma}^{\tau}\Big(\frac{1}{2}\partial_{\beta}\Gamma_{\gamma\tau}^{\sigma}+\frac{1}{3}\Gamma_{\beta\sigma}^{\tau}\Gamma_{\gamma\tau}^{\sigma}\Big)
\end{align}
where $ R\tilde{R}={R^{\mu\nu\rho}}_{\sigma} {\tilde{R}^{\sigma}}_{\;\,\rho\mu\nu}$ and $  {\tilde{R}^{\sigma}}_{\;\,\rho\mu\nu}:=\frac{1}{2}\epsilon_{\mu\nu\alpha\beta}{R^{\alpha\beta\sigma}}_{\rho} $. The current $ K^{\mu} $ vanishes in any static spacetime and does not transform covariantly. Similarly to the sGB case, this current will forcibly source scalar hair around any (non-static) black holes. One might also consider operators with higher derivatives, for instance $\phi \, \square( R_{\mu\nu\rho\sigma} R^{\mu\nu\rho\sigma})$. As remarked, for the theory to be consistent the generated hair must be small. Nonetheless, the presence of this kind of operators might be tested through future detections of gravitational waves.

\section{Conclusions}\label{sec6}
In this paper we have shown that asymptotically flat black holes in shift-symmetric scalar-tensor theories with no ghost degrees of freedom can have nontrivial scalar hair only in the presence of the operator $\phi \, \mathcal{R}^2_{GB}$ (sGB). Further assumptions include time-independence and spherical symmetry. We have laid out this fact by building from the no-hair theorem of Hui and Nicolis, which is directly applicable only to Horndeski theories. We have shown that this theorem allows a single pathology-free exception, by first addressing some concerns about the sGB solution and the infinite norm of its associated current at the black hole horizon. The fact that this object is either non diffeomorphism invariant or non-local devoids this divergence of physical meaning. Instead, any local scalar quantities were shown to be finite. In contrast, all of the other exceptions to the no-hair theorem within the realm of shift-symmetric Horndeski theories turn out to feature pathologies, such as the lack of a Lorentz invariant solution in flat space.

Stepping away from theories with second-order equations of motion, we extended the applicability of the no-hair theorem to a larger class of shift-symmetric scalar-tensor theories, which nevertheless propagate no extra degrees of freedom (the so called DHOST). Among these, we focused on those which can recover General Relativity when $X=0$, therefore selecting the class which also contains Horndeski and it is in fact generated from it by $X$-dependent invertible conformal plus disformal transformations of the metric. Leveraging this fact, we were able to show that no new operator that produces hair apart from sGB can arise in this larger class of theories, since hair cannot be generated nor removed by such transformations. Therefore, sGB remains the unquestionable champion, being the only one able to source healthy nontrivial scalar hair.

It is in the context of shift-symmetric theories in which it was ultimately possible to give a sharp answer to the question of black hole hair. This is a compelling scenario since an approximately massless scalar field can be important thoughout a large range of scales, from the cosmological to the astrophysical. One such interesting situation is when the effect of black hole hair on the production of gravitational waves in black hole mergers could help in unveiling the dynamics of the dark energy field. This scenario was put forward in \cite{Noller:2019chl}, where in spite of there being only a single possible source of hair, i.e.~sGB, the phenomenology is sensitive to the other operators present in the Lagrangian, allowing for a rich array of observational signatures.

It would be interesting to extend our study to the case of rotating BHs. For example, numerical studies have shown that regular hairy solutions with arbitrarily large rotation exist in the presence of sGB \cite{Delgado:2020rev}. However, to our knowledge, the no-hair theorem of \cite{Hui:2012qt} has only been extended to the case of slowly rotating black holes in Horndeski \cite{Sotiriou:2013qea}.

\subsection*{Acknowledgements} We thank E.~Babichev, C.~Charmousis, A.~Cisterna, M.~Crisostomi, L.~Hui, A.~Lehebel, S.~Mukohyama, A.~Nicolis, A.~Podo, T.~Sotiriou, G.~Tambalo, F.~Vernizzi, V.~Yingcharoenrat and especially M.~Mirbabayi for important discussions. ET and LGT are supported in part by the MIUR under the contract 2017FMJFMW. This research was supported by the Munich Institute for Astro- and Particle Physics (MIAPP) of the DFG cluster of excellence ``Origin and Structure of the Universe". LGT would like to thank ICTP for hospitality during part of this work.  

\appendix

\section{The many Gauss-Bonnet currents} \label{app:GBJs}
As we have discussed in Section 2, the Gauss-Bonnet invariant is the divergence of a current which is not itself a tensorial object. For this reason taking its square does not give a quantity which is invariant under diffeomorphisms.
In the main text we discussed the special expression that this current takes in the presence of a Killing vector aligned with a coordinate.  In a generic case the current can be written is terms of the spin connection.
Its expression does not give a covariant vector, in the same way the Christoffel symbols are not rank-3 tensors.

Using Greek and Latin letters for curved and flat indexes respectively, the vierbeins $ e^a_{\mu}(x) $ will be defined through the following relation: $ g_{\mu\nu}=e_{\mu}^a\eta_{ab}e_{\nu}^b $ . The spin connection will be $ {\omega^a}_b=\omega^a_{\mu b} dx^{\mu}=e^a_{\nu}\nabla_{\mu} e^{\nu}_bdx^{\mu} $ and the curvature form will be $R^{ab}=d{\omega^a}^b+{\omega^a}_c\wedge{\omega^c}^b =e^a_{\mu}e^b_{\nu}{R^{\mu\nu}}_{\rho\sigma}dx^{\rho}\wedge dx^{\sigma} $, where as usual flat indexes are lifted and lowered with the flat metric $ \eta_{ab} $. Using these definitions we can express the Gauss-Bonnet invariant as a total derivative:
\begin{align}
\int d^4x \sqrt{-g} \mathcal{R}^2_{GB}&=\int R^{ab}R^{cd}\epsilon_{abcd}=\int d\left(\epsilon_{abcd}\omega^{ab}\Big(R^{cd}-\frac{1}{3}{\omega^{c}}_{e}\omega^{ed}\Big)\!\!\right)\nonumber\\
&=-\int d^4x \sqrt{-g}\nabla_{\mu}\left(\epsilon^{\mu\nu\rho\sigma}{\epsilon_{\alpha\tau}}^{\beta\lambda}\omega^{\alpha}_{\nu\beta}\Big(\frac{1}{2}{R^{\tau}}_{\lambda\rho\sigma}-\frac{1}{3}\omega^{\tau}_{\rho\gamma}\omega^{\gamma}_{\sigma\lambda}\Big)\!\!\right)\label{truecurr}\;,
\end{align}
where $ \omega^{\alpha}_{\nu\beta}=\omega^{a}_{\nu b}e^{\alpha}_a e^{b}_{\beta} $ and $ \epsilon_{\mu\nu\rho\sigma}=e_{\mu}^a e_{\nu}^b e_{\rho}^c e_{\sigma}^d \epsilon_{abcd} $, with $ \epsilon_{abcd} $ the Levi-Civita symbol\footnote{Notice that in the analogous expression given in \cite{Sotiriou:2014pfa} the term $ -\frac{1}{3}\omega^{\tau}_{\rho\gamma}\omega^{\gamma}_{\sigma\lambda} $ was accidentally omitted.}.

Evaluating the expression \eqref{truecurr} in the Schwarzschild coordinates (with the natural induced vierbein) gives:
\begin{align}\label{Schw}
J^{\mu}_{\rm (Schw)}=\left(0,\frac{2r_s (r-2r_s)}{r^5},-\frac{4r_s \cot (\theta )}{r^5},0\right)\;,
\qquad \nabla_{\mu}J^{\mu}_{\rm (Schw)}=\frac{12r_s^2}{r^6} \;.
\end{align}
This current has a non-zero $ \theta $ component and its square diverges both at the horizon and at the poles:
\begin{align}\label{Schw2}
J_{\rm (Schw)}^2=\frac{4r_s^2 }{r^9}{\left(4 r \cot ^2(\theta )-\frac{(r-2r_s)^2}{r_s-r}\right)} \;.
\end{align}
This current is actually a linear combination of the currents defined in Section \ref{sec2}: $J^{\mu}_{\rm (Schw)} = \frac12  J_{(t)}^\mu + \frac12 J_{(\varphi)}^\mu$ (and this of course implies it has the right divergence).

The expression \eqref{truecurr} holds in any coordinate system without the need of a Killing vector. In the case of Schwarzschild space-time, in Kruskal-Szekeres coordinates $(T,R,\theta,\varphi)$, see for instance \cite{Wald:1984rg}, we have:
\be
J^{\mu}_{\rm (KS)}=\left(\frac{T \left(2r_s^2+r_s r+r^2\right)}{r_s^2 r^4},\frac{R \left(2r_s^2+r_s r+r^2\right)}{r_s^2 r^4},-\frac{4 r_s \cot (\theta )}{r^5},0\right)\;, \quad \nabla_{\mu}J^{\mu}_{\rm (KS)}=\frac{12r_s^2}{r^6}\;,
\ee
where $r$ must be understood as a function of the new coordinates $T$ and $R$.
The relation $ J_{\rm (KS)}^R/J_{\rm (KS)}^T=R/T $ implies that this current has no time component once we transform it to Schwarzschild coordinates. The radial component reads:
\begin{align}\label{JKS}
J^r_{\rm (KS \to Schw)}&=\frac{\partial r}{\partial T}J_{\rm (KS)}^T+\frac{\partial r}{\partial R}J_{\rm (KS)}^R=2\frac{\left({r-r_s}\right)}{r_sr^5}\left(2r_s^2+r_s r+r^2\right) =J^r_{\rm (Schw)}+\frac{2}{r_s r^2} \;.
\end{align}
Therefore, transforming back to Schwarzschild coordinates we obtain the current \eqref{Schw} plus a divergenceless term.
The square of the Kruskal-Szekeres current is divergent only at the poles:
\begin{align}
J^2_{\rm (KS)}=\frac{4}{r_s^2 r^9}\left[4r_s^4 r \cot ^2(\theta )+ (r-r_s) \left(2r_s^2+r_s r+r^2\right)^2\right] \;.
\end{align}
This does not coincide with Eq.~\eqref{Schw2}, as expected since $J^2$ is not a scalar.

Notice that it is possible to take an arbitrary linear combination of the various currents obtained above, and build another one with the proper divergence. For example we can combine the currents of Eqs.~\eqref{phicurrent} and \eqref{JKS}:
\be
J_{\rm (finite)}^\mu = -{J^{\mu}_{(\varphi)}} + 2 J^{\mu}_{\rm (KS \to Schw)} = \left(0, - \frac{4 r_s^2}{r^5}\left(1 - \frac{r^3}{r_s^3}\right),0,0\right) \;.
\ee
This current gives the correct divergence and has a finite norm everywhere for $r>0$:
\begin{equation}
J_{\rm (finite)}^2 = \frac{16}{r^7} (r-r_s) \left( \frac{r}{r_s} + \frac{r_s}{r} + 1 \right)^2.
\end{equation}

\section{Equivalence between sGB and Quintic Horndeski with $G_5 = \log(X)$} \label{app:H5-sGB}
                                                             
In this appendix we want to check explicitly the equivalence between the sGB operator and a shift-symmetric Quintic Horndeski with $G_5 = \log(X)$ \cite{Kobayashi:2011nu}. 

As a warm up, we can first look at the analogous case of shift-symmetric Cubic Horndeski with $G_3 = \log(X)$ in $d=2$ dimensions and the operator $\phi\,{^{(2)}}\!R$, where ${^{(2)}}\!R$ is the two-dimensional Ricci scalar (see also Ref.~\cite{Takahashi:2018yzc}). The scalar current for a generic $G_3(X)$ has the following form
\begin{equation}
 J_{H3}^\mu = G_{3X} \left( [\Pi] g^{\mu\nu} - \Pi^{\mu\nu} \right) \partial_{\nu} \phi\;.
\end{equation}
The equation of motion then reads                                                                                                                                          
\begin{eqnarray} \label{eom-H3}
\nabla_\mu J_{H3}^\mu &=& G_{3X} \left( [\Pi]^2 - [\Pi^2] \right) + 2 G_{3XX} \, \partial_\alpha \phi \partial_\beta \phi \left( [\Pi] \Pi^{\alpha \beta} -  \Pi^{\alpha \mu}\Pi_{\mu}^{\,\,\, \beta} \right)\notag \\
&&+ G_{3X}\,  g_{\alpha\beta} \, \partial_\mu \phi \, \nabla^{[\mu} \nabla^{\alpha]} \partial^\beta \phi\;.
\end{eqnarray}
Notice that the terms with three covariant derivatives acting on $\phi$ arrange in an antisymmetric way, leaving behind only a term proportional to the Riemann tensor, but no third derivatives of the field, as expected from a Horndeski Lagrangian. The above equation of motion in its current form obscures the fact that there is a choice of the function $G_3(X)$ that renders the equation $\phi$-independent (in $d=2$). In order to make this manifest, it is useful to consider the Cayley-Hamilton theorem, which states that any square matrix satisfies its own characteristic equation. In this case, consider the matrix of second derivatives of the field in a given basis, $\Pi^{\mu}_{\,\,\,\nu}$, a $d\times d$ matrix, the following local identity holds in $d=2$:  
\begin{equation}\label{CH-2d}
 (\Pi^2)^{\mu}_{\,\,\,\nu} - [\Pi] \Pi^{\mu}_{\,\,\,\nu} - \frac{1}{2}\delta^{\mu}_{\,\,\,\nu} \left( [\Pi^2] - [\Pi]^2 \right) = 0 \quad (d=2)\;.
\end{equation}
Then it is straightforward to rewrite the equation of motion \eqref{eom-H3} as follows
\begin{eqnarray} \label{eom-H3-2d}
\nabla_\mu J_{H3}^\mu \Big|_{d=2}  &=& \left( G_{3X} + X \, G_{3XX} \right) \left( [\Pi]^2 - [\Pi^2] \right) - G_{3X}\,  {}^{(2)}\!R^{\mu \nu} \partial_\mu \phi \, \partial_\nu \phi\;.  %
\end{eqnarray}
Finally, using that in $d=2$ the Ricci tensor is just ${}^{(2)}\!R_{\mu\nu} = {}^{(2)}\!R \, g_{\mu\nu}/2$ and the choice $G_3 = \log(X)$ we obtain
\begin{eqnarray} \label{eom-H3-2d-log}
\nabla_\mu J_{H3}^\mu \Big|_{d=2}  &=& - \frac{1}{2} {}^{(2)}\!R\;,  %
\end{eqnarray}
which is the expected result. 

Now let us turn to our case of interest. In what follows we are going to be more schematic, however the story is conceptually similar, but the calculations considerably more cumbersome due to the sheer amount of terms involved. A generic shift-symmetric Quintic Horndeski in $d=4$ dimensions will have a scalar current with two types of terms:
\begin{eqnarray}
 J_{H5} \sim G_{5X} \mathcal{R} (\nabla \nabla \phi) \partial \phi + G_{5XX} (\nabla \nabla \phi)^3 \partial \phi\;,
\end{eqnarray}
where $\mathcal{R}$ stands generically for the curvature. There various terms of each kind have several different contractions among the tensors, which nevertheless enjoy a particular structure due to the theory being Horndeski. The equation of motion, in turn, will schematically have the following seven types of terms,
\begin{eqnarray}
 \nabla J_{H5} &\sim& G_{5X}\Biggl[ \nabla \mathcal{R} (\nabla \nabla \phi) \partial \phi + \mathcal{R} ([\nabla, \nabla] \partial \phi) \partial \phi + \mathcal{R} (\nabla \nabla \phi)^2 \Biggr] \notag \\
 &&+ G_{5XX} \Biggl[ \mathcal{R} (\nabla \nabla \phi)^2 (\partial \phi)^2 + (\nabla \nabla \phi)^4 + (\nabla \nabla \phi)^2 ([\nabla, \nabla] \partial \phi) \partial \phi \Biggr]  \notag \\
 &&+ G_{5XXX} \, (\nabla \nabla \phi)^4 (\partial \phi)^2\;, \label{EOM-G5-schem1}
\end{eqnarray}
where the terms were arranged according to the number of $X$-derivatives acting on $G_5$. Once again notice that, since the theory is Horndeski, the equations of motion must be of second order. Indeed, the terms with $\nabla \mathcal{R}$ cancel identically by the differential Bianchi identities, while those with third derivatives acting on the scalar always appear antisymmetrically. Some of these terms are in fact the ones giving rise to terms quadratic in the curvature. We also emphasize that, much like in the cubic case, the terms contain various possible contractions. For example, in the last term of \eqref{EOM-G5-schem1} the two factors of $\partial \phi$ are not contracted to each other, and thus are not forming the combination $X$.

We rearrange once again the types of terms in the equation of motion, now in increasing powers of the curvature, obtaining, schematically,
\begin{eqnarray} \label{eom-H5}
 \nabla J_{H5} &\sim& \Biggl[ G_{5XX} (\nabla \nabla \phi)^4 + G_{5XXX} (\nabla \nabla \phi)^4 (\partial \phi)^2 \Biggr] \notag \\
 &&+ \mathcal{R} \Biggl[ G_{5X} (\nabla\nabla\phi)^2 + G_{5XX} (\nabla\nabla\phi)^2 (\partial \phi)^2 \Biggr] \notag \\
 &&+ G_{5X} \mathcal{R}^2 (\partial \phi)^2\;.
\end{eqnarray}
At this point, if one specializes to $d=4$ one can simplify the way indices are contracted so that, similarly to the Cubic example above, all the terms that have a $\partial_\alpha \phi\partial_\beta\phi$ become proportional to $X$. For the first line we make use of the  Cayley-Hamilton theorem in $d=4$. For the second line instead, it is useful to first decompose the Riemann tensor
\begin{eqnarray}\label{Ricci-decomp}
 R_{\mu\nu\rho\sigma} = C_{\mu\nu\rho\sigma} + E_{\mu\nu\rho\sigma} + S_{\mu\nu\rho\sigma}\;,
\end{eqnarray}
where $C_{\mu\nu\rho\sigma}$ is the Weyl tensor, and
\begin{eqnarray} 
 E_{\mu\nu\rho\sigma} &=& \frac{1}{d-2} \left[ g_{\mu\rho} S_{\nu\sigma} - g_{\mu\sigma} S_{\nu\rho} + g_{\nu\sigma} S_{\mu\rho} - g_{\nu\rho} S_{\mu\sigma} \right]\;, \\
 S_{\mu\nu\rho\sigma} &=& \frac{R}{d(d-1)} \left[ g_{\mu\rho} g_{\nu\sigma} - g_{\mu\sigma} g_{\nu\rho} \right]\;,
\end{eqnarray}
and $S_{\mu\nu} = R_{\mu\nu} - \frac{R}{d} g_{\mu\nu}$ is the traceless part of the Ricci tensor. The pieces involving the Ricci tensor quickly combine to be proportional to a metric. For the pieces involving the Weyl tensor, a bit more work is necessary to show this, but  it ultimately follows by exploiting the fact it is a fully traceless tensor. Once the $X$ is factorized, the resulting expression combines with the terms with one less $X$-derivative.

Finally, let us be more explicit with the part quadratic in the curvature (third line of \eqref{eom-H5}),
\begin{eqnarray}
 (\nabla_\mu J^\mu_{H5})^{(2)} = - G_{5X} \, \partial_\alpha \phi \, \partial_\beta \phi \Biggl[ R_{\mu\nu} R^{\mu\alpha\nu\beta} - \frac{1}{2} R R^{\alpha\beta} + R^{\alpha}_{\,\,\,\mu} R^{\mu\beta} - \frac{1}{2} R_{\mu\nu\rho}^{\quad\,\,\,\alpha} R^{\mu\nu\rho\beta} \Biggr]\;.
\end{eqnarray}
Using the decomposition \eqref{Ricci-decomp}, the above expression can be brought to the form
\begin{eqnarray}
  (\nabla_\mu J^\mu_{H5})^{(2)} = - G_{5X} \, \partial_\alpha \phi \, \partial_\beta \phi \Biggl[ \left( \frac{1}{4} S_{\mu\nu} S^{\mu\nu}  - \frac{R^2}{48} \right) g^{\alpha\beta} - \frac{1}{2} C_{\mu\nu\rho}^{\quad\,\,\,\alpha} C^{\mu\nu\rho\beta} \Biggr]\;,
\end{eqnarray}
where again the nontrivial part is the one involving the Weyl tensor. In this case, it is necessary to further decompose it into its electric and magnetic parts, defined as
\begin{eqnarray}
 E_{\mu\nu} = C_{\mu\alpha\nu\beta} U^{\alpha} U^{\beta}\;, \qquad
 B_{\mu\nu} = \tilde{C}_{\mu\alpha\nu\beta} U^{\alpha} U^{\beta}\;,
\end{eqnarray}
where $U^\mu$ is any timelike unit vector defining a local frame, and $\tilde{C}_{\mu\alpha\nu\beta}$ is the dual of the Weyl tensor, $  {\tilde{C}^{\sigma}}_{\;\,\rho\mu\nu}:=\frac{1}{2}\epsilon_{\mu\nu\alpha\beta}{C^{\alpha\beta\sigma}}_{\rho} $.      Here, $E_{\mu\nu}$ and $B_{\mu\nu}$ are symmetric, traceless and transverse to $U^\mu$. An explicit expression for $C_{\mu\alpha\nu\beta}$ in terms of them can be found in Ref. \cite{hansstephani2009}. With these tools, it can be shown that 
\begin{equation}
C_{\mu\nu\rho}^{\quad\,\,\,\alpha} C^{\mu\nu\rho\beta} = 2(d-4) E^{\alpha\mu} E_{\mu}^{\,\,\,\beta} + 2 (E_{\mu\nu} E^{\mu\nu} - B_{\mu\nu} B^{\mu\nu}) g^{\alpha\beta}\;.
\end{equation}
Therefore, in $d=4$, this contribution is indeed proportional to the metric. Notice that, although $E_{\mu\nu}$ and $B_{\mu\nu}$ are frame dependent, the combination on the second term above is in fact invariant. With this, we can finally write
\begin{equation}
  (\nabla_\mu J^\mu_{H5})^{(2)} = \frac{1}{8} X G_{5X} \Biggl[ -2 S_{\mu\nu} S^{\mu\nu}  + \frac{R^2}{6} + 8(E_{\mu\nu} E^{\mu\nu} - B_{\mu\nu} B^{\mu\nu}) \Biggr]\;, 
\end{equation}
the quantity in brackets being no other than the Gauss-Bonnet invariant $\mathcal{R}^2_{GB}$.

Putting everything together, the equation of motion can be written in the following form
\begin{eqnarray}
 \nabla_\mu J^\mu_{H5}\Big|_{d=4} &=& -\frac{2}{3} \left( 2G_{5XX} + X G_{5XXX} \right) \Bigl( [\Pi]^4 - 6 [\Pi^2] [\Pi]^2 + 3 [\Pi^2]^2 + 8 [\Pi^3] [\Pi] - 6 [\Pi^4] \Bigr) \notag \\
 &&-2 \left( G_{5X} + X G_{5XX} \right) \left[ \frac{1}{3} \left( [\Pi]^2 - [\Pi^2] \right) R - \left( [\Pi] \Pi_{\mu\nu} - \Pi^2_{\mu\nu} \right) R^{\mu\nu} + \Pi_{\mu\rho} \Pi_{\nu\sigma} C^{\mu\nu\rho\sigma} \right] \notag \\
 &&+ \frac{X G_{5X}}{8} \mathcal{R}^2_{GB}\;.
\end{eqnarray}
We emphasize again that we crucially rely on being in $d=4$ dimensions in order to express the equation in this form. The unique choice $G_5 = \log|X|$ makes the whole $\phi$-dependence go away, leaving only 
\begin{eqnarray}
 \nabla_\mu J^\mu_{H5} = \frac{1}{8} \mathcal{R}^2_{GB}\;. 
\end{eqnarray}

\section{Requirements on DHOST theories}
\label{app:M3}
As remarked in Section \ref{sec5}, the requirements of a ghost-free decoupling limit around flat spacetime and of the presence of the Einstein-Hilbert term define two different classes: one is generated by Horndeski Lagrangians by a conformal plus disformal transformation, while the other would appear more difficult to explore.
Here we will show that despite admitting the presence of both quadratic and cubic DHOST operators, this second class never admits a standard Einstein-Hilbert term, making it impossible to recover General Relativity in the limit in which $ X\to 0 $.

The proof makes use of the result of \cite{Achour:2016rkg}, i.e.~that every quadratic DHOST theory admitting a healthy decoupling limit is connected to the quartic Horndeski Lagrangians via an invertible conformal plus disformal transformation of the form \eqref{disf-transf}. Thus one can start by examining the condition of compatibility of the cubic part with the quadratic one, when this last is chosen to be the quartic Horndeski\footnote{The generic compatibility conditions can be found in Section 4 of \cite{BenAchour:2016fzp}, conditions (1) and (3) in the second table. These conditions become degenerate when the quadratic DHOST part is simply a quartic Horndeski theory.}:
\begin{align}\label{condition}
0=-4\frac{G_{4X}^2}{G_4}+4\frac{G_{4X}}{X}-\frac{G_4}{X^2}=-G_4\left(2\frac{G_{4X}}{G_4}-\frac{1}{X}\right)^2\;.
\end{align}
This means that either $ G_4\equiv 0 $ or $ G_4 \propto \sqrt{X} $. Both these solutions correspond to theories which contain no Einstein-Hilbert term.

If we do not restrict to quartic Horndeski, the quadratic-cubic compatibility conditions become more involved. However, knowing that all the quadratic DHOST theories that we are scanning can be obtained by a conformal plus disformal transformation of a quartic Horndeski theory, we can simply inspect how the transformation \eqref{disf-transf} will change a function $ G_4 $ that solves Eq.~\eqref{condition}:
\begin{align}
\bar{G}_{4} \sqrt{\Omega} (\Omega + X \Gamma)^{1/2}=G_{4}\propto \sqrt{X}\;.
\end{align}
This means that as long as we require $ \Omega(0)=1 $ and $ \Gamma(X) $ to be smooth in $X=0$, the function $ \bar{G_4} $ will not contain a constant term, therefore making impossible to retrieve General Relativity when $ X=0 $.
                                     
\bibliographystyle{utphys}                    
\bibliography{GB-Hair}

\providecommand{\href}[2]{#2}\begingroup\raggedright\begin{thebibliography}{10}

\bibitem{Hui:2012qt}
L.~Hui and A.~Nicolis, ``{No-Hair Theorem for the Galileon},''
  \href{http://dx.doi.org/10.1103/PhysRevLett.110.241104}{{\em Phys. Rev.
  Lett.} {\bf 110} (2013)  241104},
\href{http://arxiv.org/abs/1202.1296}{{\tt arXiv:1202.1296 [hep-th]}}.
%%CITATION = ARXIV:1202.1296;%%.

\bibitem{Sotiriou:2013qea}
T.~P. Sotiriou and S.-Y. Zhou, ``{Black hole hair in generalized scalar-tensor
  gravity},'' \href{http://dx.doi.org/10.1103/PhysRevLett.112.251102}{{\em
  Phys. Rev. Lett.} {\bf 112} (2014)  251102},
\href{http://arxiv.org/abs/1312.3622}{{\tt arXiv:1312.3622 [gr-qc]}}.
%%CITATION = ARXIV:1312.3622;%%.

\bibitem{Babichev:2017guv}
E.~Babichev, C.~Charmousis, and A.~Leh{\'e}bel, ``{Asymptotically flat black
  holes in Horndeski theory and beyond},''
  \href{http://dx.doi.org/10.1088/1475-7516/2017/04/027}{{\em JCAP} {\bf 1704}
  (2017) no.~04, 027},
\href{http://arxiv.org/abs/1702.01938}{{\tt arXiv:1702.01938 [gr-qc]}}.
%%CITATION = ARXIV:1702.01938;%%.

\bibitem{Horndeski:1974wa}
G.~W. Horndeski, ``{Second-order scalar-tensor field equations in a
  four-dimensional space},''
\href{http://dx.doi.org/10.1007/BF01807638}{{\em Int. J. Theor. Phys.} {\bf 10}
  (1974)  363--384}.
%%CITATION = IJTPB,10,363;%%.

\bibitem{Deffayet:2011gz}
C.~Deffayet, X.~Gao, D.~A. Steer, and G.~Zahariade, ``{From k-essence to
  generalised Galileons},''
  \href{http://dx.doi.org/10.1103/PhysRevD.84.064039}{{\em Phys. Rev.} {\bf
  D84} (2011)  064039},
\href{http://arxiv.org/abs/1103.3260}{{\tt arXiv:1103.3260 [hep-th]}}.
%%CITATION = ARXIV:1103.3260;%%.

\bibitem{Kobayashi:2011nu}
T.~Kobayashi, M.~Yamaguchi, and J.~Yokoyama, ``{Generalized G-inflation:
  Inflation with the most general second-order field equations},''
  \href{http://dx.doi.org/10.1143/PTP.126.511}{{\em Prog. Theor. Phys.} {\bf
  126} (2011)  511--529},
\href{http://arxiv.org/abs/1105.5723}{{\tt arXiv:1105.5723 [hep-th]}}.
%%CITATION = ARXIV:1105.5723;%%.

\bibitem{Saravani:2019xwx}
M.~Saravani and T.~P. Sotiriou, ``{Classification of shift-symmetric Horndeski
  theories and hairy black holes},''
  \href{http://dx.doi.org/10.1103/PhysRevD.99.124004}{{\em Phys. Rev.} {\bf
  D99} (2019) no.~12, 124004},
\href{http://arxiv.org/abs/1903.02055}{{\tt arXiv:1903.02055 [gr-qc]}}.
%%CITATION = ARXIV:1903.02055;%%.

\bibitem{Yagi:2015oca}
K.~Yagi, L.~C. Stein, and N.~Yunes, ``{Challenging the Presence of Scalar
  Charge and Dipolar Radiation in Binary Pulsars},''
  \href{http://dx.doi.org/10.1103/PhysRevD.93.024010}{{\em Phys. Rev.} {\bf
  D93} (2016) no.~2, 024010},
\href{http://arxiv.org/abs/1510.02152}{{\tt arXiv:1510.02152 [gr-qc]}}.
%%CITATION = ARXIV:1510.02152;%%.

\bibitem{Yale:2010jy}
A.~Yale and T.~Padmanabhan, ``{Structure of Lanczos-Lovelock Lagrangians in
  Critical Dimensions},''
  \href{http://dx.doi.org/10.1007/s10714-011-1146-1}{{\em Gen. Rel. Grav.} {\bf
  43} (2011)  1549--1570},
\href{http://arxiv.org/abs/1008.5154}{{\tt arXiv:1008.5154 [gr-qc]}}.
%%CITATION = ARXIV:1008.5154;%%.

\bibitem{Sotiriou:2014pfa}
T.~P. Sotiriou and S.-Y. Zhou, ``Black hole hair in generalized scalar-tensor
  gravity: An explicit example,''
  \href{http://dx.doi.org/10.1103/PhysRevD.90.124063}{{\em Phys.Rev.D} {\bf 90}
  (2014)  124063}, \href{http://arxiv.org/abs/1408.1698}{{\tt arXiv:1408.1698
  [gr-qc]}}.

\bibitem{Nicolis:2008in}
A.~Nicolis, R.~Rattazzi, and E.~Trincherini, ``{The Galileon as a local
  modification of gravity},''
  \href{http://dx.doi.org/10.1103/PhysRevD.79.064036}{{\em Phys. Rev.} {\bf
  D79} (2009)  064036},
\href{http://arxiv.org/abs/0811.2197}{{\tt arXiv:0811.2197 [hep-th]}}.
%%CITATION = ARXIV:0811.2197;%%.

\bibitem{Gleyzes:2014dya}
J.~Gleyzes, D.~Langlois, F.~Piazza, and F.~Vernizzi, ``{Healthy theories beyond
  Horndeski},'' \href{http://dx.doi.org/10.1103/PhysRevLett.114.211101}{{\em
  Phys. Rev. Lett.} {\bf 114} (2015) no.~21, 211101},
\href{http://arxiv.org/abs/1404.6495}{{\tt arXiv:1404.6495 [hep-th]}}.
%%CITATION = ARXIV:1404.6495;%%.

\bibitem{Gleyzes:2014qga}
J.~Gleyzes, D.~Langlois, F.~Piazza, and F.~Vernizzi, ``{Exploring gravitational
  theories beyond Horndeski},''
  \href{http://dx.doi.org/10.1088/1475-7516/2015/02/018}{{\em JCAP} {\bf 1502}
  (2015)  018},
\href{http://arxiv.org/abs/1408.1952}{{\tt arXiv:1408.1952 [astro-ph.CO]}}.
%%CITATION = ARXIV:1408.1952;%%.

\bibitem{Langlois:2015cwa}
D.~Langlois and K.~Noui, ``{Degenerate higher derivative theories beyond
  Horndeski: evading the Ostrogradski instability},''
  \href{http://dx.doi.org/10.1088/1475-7516/2016/02/034}{{\em JCAP} {\bf 1602}
  (2016) no.~02, 034},
\href{http://arxiv.org/abs/1510.06930}{{\tt arXiv:1510.06930 [gr-qc]}}.
%%CITATION = ARXIV:1510.06930;%%.

\bibitem{Crisostomi:2016tcp}
M.~Crisostomi, M.~Hull, K.~Koyama, and G.~Tasinato, ``{Horndeski: beyond, or
  not beyond?},'' \href{http://dx.doi.org/10.1088/1475-7516/2016/03/038}{{\em
  JCAP} {\bf 1603} (2016) no.~03, 038},
\href{http://arxiv.org/abs/1601.04658}{{\tt arXiv:1601.04658 [hep-th]}}.
%%CITATION = ARXIV:1601.04658;%%.

\bibitem{Achour:2016rkg}
J.~Ben~Achour, D.~Langlois, and K.~Noui, ``{Degenerate higher order
  scalar-tensor theories beyond Horndeski and disformal transformations},''
  \href{http://dx.doi.org/10.1103/PhysRevD.93.124005}{{\em Phys. Rev.} {\bf
  D93} (2016) no.~12, 124005},
\href{http://arxiv.org/abs/1602.08398}{{\tt arXiv:1602.08398 [gr-qc]}}.
%%CITATION = ARXIV:1602.08398;%%.

\bibitem{BenAchour:2016fzp}
J.~Ben~Achour, M.~Crisostomi, K.~Koyama, D.~Langlois, K.~Noui, and G.~Tasinato,
  ``{Degenerate higher order scalar-tensor theories beyond Horndeski up to
  cubic order},'' \href{http://dx.doi.org/10.1007/JHEP12(2016)100}{{\em JHEP}
  {\bf 12} (2016)  100},
\href{http://arxiv.org/abs/1608.08135}{{\tt arXiv:1608.08135 [hep-th]}}.
%%CITATION = ARXIV:1608.08135;%%.

\bibitem{BenAchour:2019fdf}
J.~Ben~Achour, H.~Liu, and S.~Mukohyama, ``{Hairy black holes in DHOST
  theories: Exploring disformal transformation as a solution-generating
  method},'' \href{http://dx.doi.org/10.1088/1475-7516/2020/02/023}{{\em JCAP}
  {\bf 2002} (2020) no.~02, 023},
\href{http://arxiv.org/abs/1910.11017}{{\tt arXiv:1910.11017 [gr-qc]}}.
%%CITATION = ARXIV:1910.11017;%%.

\bibitem{Ripley:2019hxt}
J.~L. Ripley and F.~Pretorius, ``{Hyperbolicity in Spherical Gravitational
  Collapse in a Horndeski Theory},''
  \href{http://dx.doi.org/10.1103/PhysRevD.99.084014}{{\em Phys. Rev.} {\bf
  D99} (2019) no.~8, 084014},
\href{http://arxiv.org/abs/1902.01468}{{\tt arXiv:1902.01468 [gr-qc]}}.
%%CITATION = ARXIV:1902.01468;%%.

\bibitem{Okounkova:2017yby}
M.~Okounkova, L.~C. Stein, M.~A. Scheel, and D.~A. Hemberger, ``{Numerical
  binary black hole mergers in dynamical Chern-Simons gravity: Scalar field},''
  \href{http://dx.doi.org/10.1103/PhysRevD.96.044020}{{\em Phys. Rev.} {\bf
  D96} (2017) no.~4, 044020},
\href{http://arxiv.org/abs/1705.07924}{{\tt arXiv:1705.07924 [gr-qc]}}.
%%CITATION = ARXIV:1705.07924;%%.

\bibitem{Noller:2019chl}
J.~Noller, L.~Santoni, E.~Trincherini, and L.~G. Trombetta, ``{Black Hole
  Ringdown as a Probe for Dark Energy},''
\href{http://arxiv.org/abs/1911.11671}{{\tt arXiv:1911.11671 [gr-qc]}}.
%%CITATION = ARXIV:1911.11671;%%.

\bibitem{Delgado:2020rev}
J.~F.~M. Delgado, C.~A.~R. Herdeiro, and E.~Radu, ``{Spinning black holes in
  shift-symmetric Horndeski theory},''
\href{http://arxiv.org/abs/2002.05012}{{\tt arXiv:2002.05012 [gr-qc]}}.
%%CITATION = ARXIV:2002.05012;%%.

\bibitem{Wald:1984rg}
R.~M. Wald,
  \href{http://dx.doi.org/10.7208/chicago/9780226870373.001.0001}{{\em {General
  Relativity}}}.
\newblock Chicago Univ. Pr., Chicago, USA,
1984.
\newblock
%%CITATION = INSPIRE-209356;%%.

\bibitem{Takahashi:2018yzc}
K.~Takahashi and T.~Kobayashi, ``{Generalized 2D dilaton gravity and kinetic
  gravity braiding},'' \href{http://dx.doi.org/10.1088/1361-6382/ab1355}{{\em
  Class. Quant. Grav.} {\bf 36} (2019) no.~9, 095003},
\href{http://arxiv.org/abs/1812.08847}{{\tt arXiv:1812.08847 [gr-qc]}}.
%%CITATION = ARXIV:1812.08847;%%.

\bibitem{hansstephani2009}
H.~Stephani, {\em Exact Solutions of Einstein's Field Equations (Cambridge
  Monographs on Mathematical Physics)}.
\newblock Cambridge University Press, oct, 2009.
\newblock \url{https://www.xarg.org/ref/a/0521467020/}.

\end{thebibliography}\endgroup

\end{document}